\documentclass[aps,twocolumn,floatfix,eqsecnum,showpacs,prb]{revtex4}
\usepackage{graphicx}
\usepackage{amsmath}
\usepackage{amssymb}
\usepackage{bm}
\usepackage{color}

\begin{document}

\title{Nernst effect beyond the relaxation-time approximation}
\author{D. I. Pikulin}
\affiliation{Instituut-Lorentz, Universiteit Leiden, P.O. Box 9506, 2300 RA Leiden, The Netherlands}
\author{C.-Y. Hou}
\affiliation{Instituut-Lorentz, Universiteit Leiden, P.O. Box 9506, 2300 RA Leiden, The Netherlands}
\author{C. W. J. Beenakker}
\affiliation{Instituut-Lorentz, Universiteit Leiden, P.O. Box 9506, 2300 RA Leiden, The Netherlands}

\date{May, 2011}

\begin{abstract}
Motivated by recent interest in the Nernst effect in cuprate superconductors, we calculate this magneto-thermo-electric effect for an arbitrary (anisotropic) quasiparticle dispersion relation and elastic scattering rate. The exact solution of the linearized Boltzmann equation is compared with the commonly used relaxation-time approximation. We find qualitative deficiencies of this approximation, to the extent that it can get the sign wrong of the Nernst coefficient. Ziman's improvement of the relaxation-time approximation, which becomes exact when the Fermi surface is isotropic, also cannot capture the combined effects of anisotropy in dispersion and scattering.
\end{abstract}

\maketitle

\section{Introduction}
\label{intro}

The Nernst effect is a magneto-thermo-electric effect, in which an electric field $E_{x}$ in the $x$-direction results from a temperature gradient $\partial T/\partial y$ in the $y$-direction, in the presence of a (weak) magnetic field $B$ in the $z$-direction.\cite{Del65} The Nernst coefficient ${\cal N}_{xy}=-E_{x}(B\partial T/\partial y)^{-1}$ depends sensitively on anisotropies in the band structure. In particular, for a square lattice ${\cal N}_{xy}=-{\cal N}_{yx}$ is antisymmetric upon interchange of $x$ and $y$ --- just like the Hall resistivity --- but lattice distortion breaks this antisymmetry.

There has been much recent interest in the Nernst effect in the context of high-$T_{c}$ superconductivity, since underdoped cuprates were found to have an unusually large Nernst coefficient in the normal state.\cite{Xu00} This may be due to superconducting fluctuations above $T_{c}$, or it may be purely a quasiparticle effect.\cite{Beh09} The quasiparticle Nernst effect has been studied on the basis of the linearized Boltzmann equation in the relaxation-time approximation.\cite{Lam96,Oga04,Hac09a,Hac09b,Hac10a,Hac10b,Zha10} This is a reliable approach if the scattering rate is isotropic, since then the neglected ``scattering-in'' contributions average out to zero. There is, however, considerable experimental evidence for predominantly small-angle elastic scattering in the cuprates,\cite{Val00,Kam05,Cha08,Nar08} possibly due to long-range potential fluctuations from dopant atoms in between the ${\rm CuO_{2}}$ planes.\cite{Abr00,Zhu04}

It is not surprising that existing studies rely on the relaxation-time approximation, since the full solution of the Boltzmann equation with both band and scattering anisotropies is a notoriously difficult problem.\cite{Zim72} In our literature search we have found magneto-electric calculations that go beyond the relaxation-time approximation,\cite{Jon69,Hlu01,Car02,Smi08} but no magneto-thermo-electric studies. It is the purpose of this paper to provide such a calculation and to assess the reliability of the relaxation-time approximation.

We start in Sec.\ \ref{transportproblem} with a formulation of the anisotropic transport problem, in terms of the socalled vector mean free path.\cite{Son62,Tay63} In the relaxation-time approximation, this vector $\bm{\Lambda}_{\bm{k}}$ is simply given by the product ${\bm v}_{\bm{k}}\tau_{\bm{k}}$ of velocity and scattering time (all quantities dependent on the point $\bm{k}$ on the Fermi surface). Going beyond this approximation, $\bm{\Lambda}_{\bm{k}}$ is determined by an integral equation, which we solve numerically. 

We also consider, in Sec.\ \ref{relaxationtime}, an improvement on the relaxation-time approximation, due to Ziman,\cite{Zim72,Zim61} which incorporates some of the scattering-in contributions into the definition of the scattering time. For isotropic Fermi surfaces Ziman's scattering time is just the familiar transport mean free time --- which fully accounts for scattering anisotropies. If the dispersion relation is not isotropic this is no longer the case.  

We compare the exact and approximate solutions in Sec.\ \ref{compare} and conclude in Sec.\ \ref{conclude}.

\section{Formulation of the transport problem}
\label{transportproblem}

\subsection{Boltzmann equation}
\label{sec_Boltzmann}

We start from the semiclassical Boltzmann transport equation for quasiparticles (charge $e$) in a weak magnetic field $\bm{B}$, driven out of equilibrium by a spatially uniform electric field $\bm{E}$ and temperature gradient $\nabla T$. The excitation energy is $\varepsilon_{\bm{k}}$, relative to the Fermi energy $\varepsilon_{F}$. The band structure may be anisotropic, so that the velocity
\begin{equation}
\bm{v}_{\bm{k}}=\hbar^{-1}\nabla_{\bm{k}}\varepsilon_{\bm{k}}\label{vdef}
\end{equation}
(with $\nabla_{\bm{k}}=\partial/\partial\bm{k}$) need not be parallel to the momentum $\hbar\bm{k}$. For simplicity, we assume there is only a single type of carriers at the Fermi level (either electrons or holes).

Upon linearization of the distribution function $f_{\bm{k}}=f_{0}+g_{\bm{k}}$ around the equilibrium solution
\begin{equation}
f_{0}=\frac{1}{1+\exp[(\varepsilon_{\bm{k}}-\varepsilon_{F})/k_{B}T]},\label{f0def} 
\end{equation}
the Boltzmann equation takes the form\cite{Zim72}
\begin{align}
&\bm{v}_{\bm{k}}\cdot\bm{U}-\frac{e}{\hbar} (\bm{v}_{\bm{k}} \times \bm{B})\cdot \nabla_{\bm{k}} g_{\bm{k}} = \sum_{\bm{k}'} Q(\bm{k},\bm{k}')(g_{\bm{k}}-g_{\bm{k}'}),\label{Boltzmann}\\
&\bm{U}=\left(e\bm{E}-\frac{\varepsilon_{\bm{k}} - \varepsilon_{F}}{T}\nabla T\right)\left(-\frac{\partial f_{0}} {\partial \varepsilon_{\bm{k}}}\right).
\label{Boltzmannb}
\end{align}
The right-hand-side of Eq.\ \eqref{Boltzmann} is the difference between the scattering-in term $\sum_{\bm{k}'}Q(\bm{k},\bm{k}')g_{\bm{k}'}$ and the scattering-out term $\sum_{\bm{k}'}Q(\bm{k}',\bm{k})g_{\bm{k}}$ (with $Q(\bm{k}',\bm{k})=Q(\bm{k},\bm{k}')$ because of detailed balance).

We assume elastic scattering with rate
\begin{equation}
Q(\bm{k},\bm{k}')=\delta(\varepsilon_{\bm{k}}-\varepsilon_{\bm{k}'})q(\bm{k},\bm{k}')\label{Qqdef}
\end{equation}
from $\bm{k}'$ to $\bm{k}$. Detailed balance requires
\begin{equation}
q(\bm{k}',\bm{k})=q(\bm{k},\bm{k}')\label{qsymm}
\end{equation}
and particle conservation requires
\begin{equation}
\sum_{\bm{k}} g_{\bm{k}}=0.\label{sumkgk}
\end{equation}
The sum over $\bm{k}$ represents a $d$-dimensional momentum integral, $\sum_{\bm{k}}\rightarrow (2\pi)^{-d}\int d\bm{k}$ (in a unit volume). The spin degree of freedom is omitted.

It is convenient to define the Fermi surface average
\begin{equation}
\langle f(\bm{k})\rangle_{S_{F}}=\frac{\oint dS_{F}\,f(\bm{k})|\bm{v}_{\bm{k}}|^{-1}}{\oint dS_{F}\,|\bm{v}_{\bm{k}}|^{-1}},\label{Fermiaverage}
\end{equation}
with a weight factor $|v_{\bm{k}}|^{-1}$ from the volume element $d\bm{k}=\hbar^{-1}|v_{\bm{k}}|^{-1}d\varepsilon_{\bm{k}}dS_{F}$. The density of states is given by
\begin{equation}
N(\varepsilon_{F})=\hbar^{-1}(2\pi)^{-d}\oint dS_{F}\,|\bm{v}_{\bm{k}}|^{-1}.\label{NEFdef}
\end{equation}
For later use we note the identity
\begin{equation}
\langle f(\bm{k})(\bm{v}_{\bm{k}}\times\nabla_{\bm{k}})g(\bm{k})\rangle_{S_{F}}=-\langle g(\bm{k})(\bm{v}_{\bm{k}}\times\nabla_{\bm{k}})f(\bm{k})\rangle_{S_{F}},\label{partialintSF}
\end{equation}
valid for arbitrary functions $f,g$ of $\bm{k}$.

\subsection{Vector mean free paths}
\label{vectormfp}

We seek the solution of Eq.\ \eqref{Boltzmann} to first order in $B$. Following Refs.\ \onlinecite{Son62,Tay63} we introduce the vector mean free paths $\bm{\Lambda}_{\bm{k}}$ (of order $B^{0}$) and $\delta\bm{\Lambda}_{\bm{k}}$ (of order $B^{1}$), by substituting
\begin{equation}
g_{\bm{k}}= \bm{U}\cdot\left(\bm{\Lambda}_{\bm{k}}+\delta\bm{\Lambda}_{\bm{k}}\right).\label{Substitution}
\end{equation}
Since the vector $\bm{U}$ can have an arbitrary direction it cancels from the equation for $\bm{\Lambda}_{\bm{k}}$. The equation for $\delta\bm{\Lambda}_{\bm{k}}$ has also a term $\propto(\bm{v}_{\bm{k}}\times\nabla_{\bm{k}})\bm{U}$, which vanishes because $\nabla_{\bm{k}}\bm{U}=\hbar\bm{v}_{\bm{k}}\partial\bm{U}/\partial\varepsilon_{\bm{k}}$. 

The resulting equations for the vector mean free paths are
\begin{align}
&\sum_{\bm{k}'} Q(\bm{k},\bm{k}')(\bm{\Lambda}_{\bm{k}} - \bm{\Lambda}_{\bm{k}'}) = \bm{v}_{\bm{k}}, \label{Lambda}\\
&\sum_{\bm{k}'} Q(\bm{k},\bm{k}')(\delta\bm{\Lambda}_{\bm{k}} - \delta\bm{\Lambda}_{\bm{k}'})=\frac{e}{\hbar}\bm{B}\cdot (\bm{v}_{\bm{k}}\times\nabla_{\bm{k}}) \bm{\Lambda}_{\bm{k}}. \label{deltaLambda}
\end{align}
They can be written in terms of Fermi surface averages,
\begin{align}
&N(\varepsilon_{F})\langle q(\bm{k},\bm{k}')(\bm{\Lambda}_{\bm{k}} - \bm{\Lambda}_{\bm{k}'})\rangle_{S'_{F}}=\bm{v}_{\bm{k}}, \label{LambdaF}\\
&N(\varepsilon_{F})\langle q(\bm{k},\bm{k}')(\delta\bm{\Lambda}_{\bm{k}} - \delta\bm{\Lambda}_{\bm{k}'})\rangle_{S'_{F}}=\frac{e}{\hbar}\bm{B}\cdot (\bm{v}_{\bm{k}}\times\nabla_{\bm{k}}) \bm{\Lambda}_{\bm{k}}. \label{deltaLambdaF}
\end{align}
(The prime in the subscript $S'_{F}$ indicates that $\bm{k}'$ is averaged over the Fermi surface, at fixed $\bm{k}$.) The solution should satisfy the normalization
\begin{equation}
\langle\bm{\Lambda}_{\bm{k}}\rangle_{S_{F}}=0=\langle\bm{\delta\Lambda}_{\bm{k}}\rangle_{S_{F}},\label{Lambdanorm}
\end{equation}
required by particle conservation to each order in $B$.

The integral equations \eqref{Lambda} and \eqref{deltaLambda} can be readily solved numerically. In the limit of small-angle scattering an analytical solution is possible, by expanding the $\bm{k}'$-dependence around $\bm{k}$ to second order,\cite{Hlu00,Dah10} but we have not pursued that method here.

\subsection{Linear response coefficients}
\label{transportcoeff}

In linear response the electric current density $\bm{j}$ is related to the electric field $\bm{E}$ and temperature gradient $\nabla T$ by 
\begin{equation}
\bm{j}=\bm{\sigma}\bm{E}-\bm{\alpha}\nabla T.\label{linearresponse}
\end{equation}
The conductivity tensor $\bm{\sigma}$ follows from the vector mean free paths by
\begin{align}
\bm{\sigma}&=\sum_{\bm{k}}e\bm{v}_{\bm{k}} \otimes\frac{\partial g_{\bm{k}}}{\partial\bm{E}} \nonumber\\
&=e^{2}\sum_{\bm{k}}\left(-\frac{\partial f_{0}}{\partial \varepsilon_{\bm{k}}}\right) \bm{v}_{\bm{k}} \otimes 
\left(\bm{\Lambda}_{\bm{k}}+\delta \bm{\Lambda}_{\bm{k}}\right). \label{sigma_Lambda}
\end{align}
[The direct product indicates a dyadic tensor with elements $(\bm{a}\otimes\bm{b})_{ij}=a_{i}b_{j}$.] 

At low temperatures, when $-\partial f_{0}/\partial\varepsilon_{\bm{k}}\rightarrow\delta(\varepsilon_{\bm{k}}-\varepsilon_{F})$, this may also be written as a Fermi surface average,
\begin{equation}
\bm{\sigma}=e^{2}N(\varepsilon_{F})\langle\bm{v}_{\bm{k}}\otimes(\bm{\Lambda}_{\bm{k}}+\delta\bm{\Lambda}_{\bm{k}})\rangle_{S_{F}}.\label{sigmaFermiaverage}
\end{equation}
By substituting Eq.\ \eqref{LambdaF} for $\bm{v}_{\bm{k}}$ and using Eq.\ \eqref{deltaLambdaF} together with the detailed balance condition \eqref{qsymm} and the identity \eqref{partialintSF}, one verifies the Onsager reciprocity relation
\begin{equation}
\sigma_{ij}(\bm{B})=\sigma_{ji}(-\bm{B}).\label{Onsagersymm}
\end{equation}

The thermoelectric tensor $\bm{\alpha}$ is given by
\begin{align}
\bm{\alpha}&=\sum_{\bm{k}}e\bm{v}_{\bm{k}} \otimes\frac{\partial g_{\bm{k}}}{\partial(-\nabla T)} \nonumber\\
&=\frac{e}{T}\sum_{\bm{k}} (\varepsilon_{\bm{k}}-\varepsilon_{F})\left(-\frac{\partial f_{0}}{\partial \varepsilon_{\bm{k}}}\right) \bm{v}_{\bm{k}} \otimes 
\left(\bm{\Lambda}_{\bm{k}}+\delta \bm{\Lambda}_{\bm{k}}\right). \label{alpha_Lambda}
\end{align}
At low temperatures this reduces to the Mott formula,
\begin{equation}
\bm{\alpha}=-\frac{\pi^{2}k_{B}^{2}T}{3e}\frac{d}{d\varepsilon_{F}}\bm{\sigma}.\label{Mott}
\end{equation}

These equations all refer to a single type of carriers at the Fermi level (electrons or holes), as would be appropriate for hole-doped cuprates. The ambipolar effects of coexisting electron and hole bands are not considered here. 

\subsection{Nernst effect}
\label{secNernst}

We take a two-dimensional ($d=2$) layered geometry in the $x-y$ plane, with a magnetic field $\bm{B}=B\hat{z}$ in the $z$-direction. The Nernst effect relates a transverse electric field, say in the $x$-direction, to a longitudinal temperature gradient (in the $y$-direction), for zero electric current. 

One distinguishes the isothermal and adiabatic Nernst effect,\cite{Del65} depending on whether $\partial T/\partial x=0$ or $j_{h,x}=0$ is enforced (with $\bm{j}_{h}$ the heat current). As is appropriate for the cuprates,\cite{Wan01} we assume that a high phonon contribution to the thermal conductivity keeps the transverse temperature gradient $\partial T/\partial x$ negligibly small, so that the Nernst effect is measured under isothermal conditions.

The isothermal Nernst effect is expressed by
\begin{equation}
E_{x}=\theta_{xy}\left(-\frac{\partial T}{\partial y}\right),\;\;\frac{\partial T}{\partial x}=0,\;\;\bm{j}_{e}=0,\label{ExdTdyiso}
\end{equation}
and similarly with $x$ and $y$ interchanged. The thermopower tensor
\begin{equation}
\bm{\theta}=-\bm{\sigma}^{-1}\bm{\alpha}\label{thetadef}
\end{equation}
has off-diagonal elements
\begin{subequations}
\label{Nernst}
\begin{align}
&\theta_{xy}=-\frac{\sigma_{yy}\alpha_{xy}-\sigma_{xy}\alpha_{yy}}{\sigma_{xx}\sigma_{yy}-\sigma_{xy}\sigma_{yx}},\label{Nernsta}\\
&\theta_{yx}=-\frac{\sigma_{xx}\alpha_{yx}-\sigma_{yx}\alpha_{xx}}{\sigma_{xx}\sigma_{yy}-\sigma_{xy}\sigma_{yx}}.\label{Nernstb}
\end{align}
\end{subequations}

We will consider two-dimensional anisotropic band structures that still possess at least one axis of reflection symmetry, say the $y$-axis. Upon reflection the component $j_{x}\mapsto-j_{x}$ of the electric current changes sign, while $E_{y}$ and $\partial T/\partial y$ remain unchanged. The perpendicular magnetic field $B\mapsto -B$ also changes sign, because it is an axial vector. It follows that $\sigma_{xy}(B)=-\sigma_{xy}(-B)$ and $\alpha_{xy}(B)=-\alpha_{xy}(-B)$ are both odd functions of $B$, so they vanish when $B\rightarrow 0$. 

Using the Mott formula \eqref{Mott}, one can then define the $B$-independent Nernst coefficients
\begin{subequations}
\label{calNdef}
\begin{align}
{\cal N}_{xy}&=\lim_{B\rightarrow 0}\theta_{xy}/B\nonumber\\
&=\frac{\pi^{2}k_{B}^{2}T}{3e}\lim_{B\rightarrow 0}\frac{1}{B\sigma_{xx}}\left(\frac{d\sigma_{xy}}{d\varepsilon_{F}}-\frac{\sigma_{xy}}{\sigma_{yy}}\frac{d\sigma_{yy}}{d\varepsilon_{F}}\right)\nonumber\\
&=\frac{\pi^{2}k_{B}^{2}T}{3e}\lim_{B\rightarrow 0}\frac{1}{B}\frac{\sigma_{yy}}{\sigma_{xx}}\frac{d}{d\varepsilon_{F}}\frac{\sigma_{xy}}{\sigma_{yy}},\label{calNdefa}\\
{\cal N}_{yx}&=-\frac{\pi^{2}k_{B}^{2}T}{3e}\lim_{B\rightarrow 0}\frac{1}{B}\frac{\sigma_{xx}}{\sigma_{yy}}\frac{d}{d\varepsilon_{F}}\frac{\sigma_{xy}}{\sigma_{xx}}.\label{calNdefb}
\end{align}
\end{subequations}
These expressions relate the Nernst coefficients to the energy derivative of the Hall angle in the small magnetic-field limit. The cancellation in Eq.\ \eqref{calNdefa} of any identical energy dependence of $\sigma_{xy}$ and $\sigma_{yy}$ is known as the Sondheimer cancellation.\cite{Beh09,Son48} On a square lattice one has $\sigma_{xx}=\sigma_{yy}$, hence ${\cal N}_{xy}=-{\cal N}_{yx}$, but without this $C_{4}$ symmetry the two Nernst coefficients differ in absolute value.

In terms of the vector mean free paths, the Nernst coefficients are given by
\begin{subequations}
\label{calNLambda}
\begin{align}
{\cal N}_{xy}&=\frac{\pi^{2}k_{B}^{2}T}{3eB}\frac{\langle v_{\bm{k},y}\Lambda_{\bm{k},y}\rangle_{S_{F}}}{\langle v_{\bm{k},x}\Lambda_{\bm{k},x}\rangle_{S_{F}}}\frac{d}{d\varepsilon_{F}}\frac{\langle v_{\bm{k},x}\delta\Lambda_{\bm{k},y}\rangle_{S_{F}}}{\langle v_{\bm{k},y}\Lambda_{\bm{k},y}\rangle_{S_{F}}},\label{calNLambdaa}\\
{\cal N}_{yx}&=-\frac{\pi^{2}k_{B}^{2}T}{3eB}\frac{\langle v_{\bm{k},x}\Lambda_{\bm{k},x}\rangle_{S_{F}}}{\langle v_{\bm{k},y}\Lambda_{\bm{k},y}\rangle_{S_{F}}}\frac{d}{d\varepsilon_{F}}\frac{\langle v_{\bm{k},x}\delta\Lambda_{\bm{k},y}\rangle_{S_{F}}}{\langle v_{\bm{k},x}\Lambda_{\bm{k},x}\rangle_{S_{F}}},\label{calNLambdab}
\end{align}
\end{subequations}
where we have used that $\bm{\Lambda}_{\bm{k}}$ is $B$-independent and $\delta\bm{\Lambda}_{\bm{k}}$ is $\propto B$.

\section{Relaxation-time approximation}
\label{relaxationtime}

In the relaxation-time approximation the scattering-in term $\sum_{\bm{k}'}Q(\bm{k},\bm{k}')g_{\bm{k}'}$ on the right-hand-side of the Boltzmann equation \eqref{Boltzmann} is omitted.\cite{Zim72} Only the scattering-out term $g_{\bm{k}}\sum_{\bm{k}'}Q(\bm{k},\bm{k}')=g_{\bm{k}}/\tau_{\bm{k}}$ is retained, containing the momentum dependent relaxation rate
\begin{equation}
1/\tau_{\bm{k}}=\sum_{\bm{k}'}Q(\bm{k},\bm{k}')=N(\varepsilon_{F})\langle q(\bm{k},\bm{k}')\rangle_{S'_{F}}.\label{taukdef}
\end{equation}

Without the scattering-in term, the equations \eqref{Lambda} and \eqref{deltaLambda} for the vector mean free paths can be solved immediately,
\begin{equation}
\bm{\Lambda}_{\bm{k}}=\bm{v}_{\bm{k}}\tau_{\bm{k}},\;\;
\delta\bm{\Lambda}_{\bm{k}}=\frac{e}{\hbar}\tau_{\bm{k}}\bm{B}\cdot(\bm{v}_{\bm{k}}\times\nabla_{\bm{k}})\tau_{\bm{k}}\bm{v}_{\bm{k}}.\label{Lambdatau}
\end{equation}
In general this solution does not satisfy the particle conservation requirement \eqref{Lambdanorm}, which is the fundamental deficiency of the relaxation-time approximation.

Substitution into Eq.\ \eqref{sigmaFermiaverage} gives the conductivity tensor
\begin{equation}
\bm{\sigma}=e^{2}N(\varepsilon_{F})\langle\tau_{\bm{k}}\bm{v}_{\bm{k}}\otimes(\bm{v}_{\bm{k}}+\Omega_{\bm{k}}\tau_{\bm{k}}\bm{v}_{\bm{k}})\rangle_{S_{F}},\label{sigmaFermiaveragereltime}
\end{equation}
with differential operator
\begin{equation}
\Omega_{\bm{k}}=\frac{e}{\hbar}\bm{B}\cdot(\bm{v}_{\bm{k}}\times\nabla_{\bm{k}}).\label{Omegadef}
\end{equation}

For a two-dimensional lattice with reflection symmetry in the $y$-axis, the elements of the conductivity tensor are given by
\begin{align}
&\sigma_{xx}=e^{2}N(\varepsilon_{F})\langle\tau v_{x}^{2}\rangle_{S_{F}},\;\;\sigma_{yy}=e^{2}N(\varepsilon_{F})\langle\tau v_{y}^{2}\rangle_{S_{F}},\label{sigmaxx}\\
&\sigma_{xy}=-\sigma_{yx}=e^{2}N(\varepsilon_{F})\frac{eB}{\hbar}\nonumber\\
&\quad\times\left\langle\tau v_{x}\left(v_{x}\frac{\partial}{\partial k_{y}}-v_{y}\frac{\partial}{\partial k_{x}}\right)\tau v_{y}\right\rangle_{S_{F}}.\label{sigmaxy}
\end{align}
(Here we don't write the subscript $\bm{k}$ to simplify the notation.) The Nernst coefficients in the relaxation-time approximation then follow from Eq.\ \eqref{calNdef} as the energy derivative of the ratio of two Fermi surface averages,
\begin{subequations}
\label{calNdefreltime}
\begin{align}
{\cal N}_{xy}&=Z_{0}\frac{\sigma_{yy}}{\sigma_{xx}}\frac{d}{d\varepsilon_{F}}\frac{
\left\langle\tau v_{x}\left(v_{x}\frac{\partial}{\partial k_{y}}-v_{y}\frac{\partial}{\partial k_{x}}\right)\tau v_{y}\right\rangle_{S_{F}}}
{\langle\tau v_{y}^{2}\rangle_{S_{F}}},\label{calNdefreltimea}\\
{\cal N}_{yx}&=-Z_{0}\frac{\sigma_{xx}}{\sigma_{yy}}\frac{d}{d\varepsilon_{F}}\frac{
\left\langle\tau v_{x}\left(v_{x}\frac{\partial}{\partial k_{y}}-v_{y}\frac{\partial}{\partial k_{x}}\right)\tau v_{y}\right\rangle_{S_{F}}}
{\langle\tau v_{x}^{2}\rangle_{S_{F}}},\label{calNdefreltimeb}
\end{align}
\end{subequations}
where we have defined
\begin{equation}
Z_{0}=\frac{\pi^{2}k_{B}^{2}T}{3\hbar}.\label{N0gammadef}
\end{equation}

One may further simplify the relaxation-time approximation by taking an isotropic relaxation time $\tau_{0}(\varepsilon_{F})$, which is the approach taken in Refs.\ \onlinecite{Hac09a,Hac09b,Hac10a,Hac10b,Zha10}. Since $(\bm{v}_{\bm{k}}\times\nabla_{\bm{k}})\tau_{0}(\varepsilon_{F})=0$, Eq.\ \eqref{calNdefreltime} then reduces to
\begin{subequations}
\label{calNdefreltime0}
\begin{align}
{\cal N}_{xy}&=Z_{0}\frac{\sigma_{yy}}{\sigma_{xx}}\frac{d}{d\varepsilon_{F}}\frac{\tau_{0}(\varepsilon_{F})}{\langle v_{y}^{2}\rangle_{S_{F}}}
\left\langle v_{x}^{2}\frac{\partial v_{y}}{\partial k_{y}}-v_{x}v_{y}\frac{\partial v_{y}}{\partial k_{x}}\right\rangle_{S_{F}},\label{calNdefreltime0a}\\
{\cal N}_{yx}&=-Z_{0}\frac{\sigma_{xx}}{\sigma_{yy}}\frac{d}{d\varepsilon_{F}}\frac{\tau_{0}(\varepsilon_{F})}{\langle v_{x}^{2}\rangle_{S_{F}}}
\left\langle v_{x}^{2}\frac{\partial v_{y}}{\partial k_{y}}-v_{x}v_{y}\frac{\partial v_{y}}{\partial k_{x}}\right\rangle_{S_{F}}.\label{calNdefreltime0b}
\end{align}
\end{subequations}

If one stays with a momentum dependent relaxation time $\tau_{\bm{k}}$, then it is possible to improve on the relaxation-time approximation by changing the definition \eqref{taukdef} into Ziman's expression\cite{Zim72,Zim61}
\begin{equation}
1/\tau_{\bm{k}}^{\rm Ziman}=N(\varepsilon_{F})\left\langle q(\bm{k},\bm{k}')\left(1-\frac{\bm{v}_{\bm{k}}\cdot\bm{v}_{\bm{k}'}}{|\bm{v}_{\bm{k}}|\,|\bm{v}_{\bm{k}'}|}\right)\right\rangle_{S'_{F}}.\label{taukdefZiman}
\end{equation}
Ziman's improvement of the relaxation-time approximation becomes exact if the Fermi surface is isotropic, meaning that $\varepsilon_{\bm{k}}$ is only a function of $|\bm{k}|$ and $q(\bm{k},\bm{k}')$ is only a function of $\bm{k}\cdot\bm{k}'$.
  
\section{Comparison}
\label{compare}

\begin{figure}[tbh]
\centerline{\includegraphics[width=0.9\linewidth]{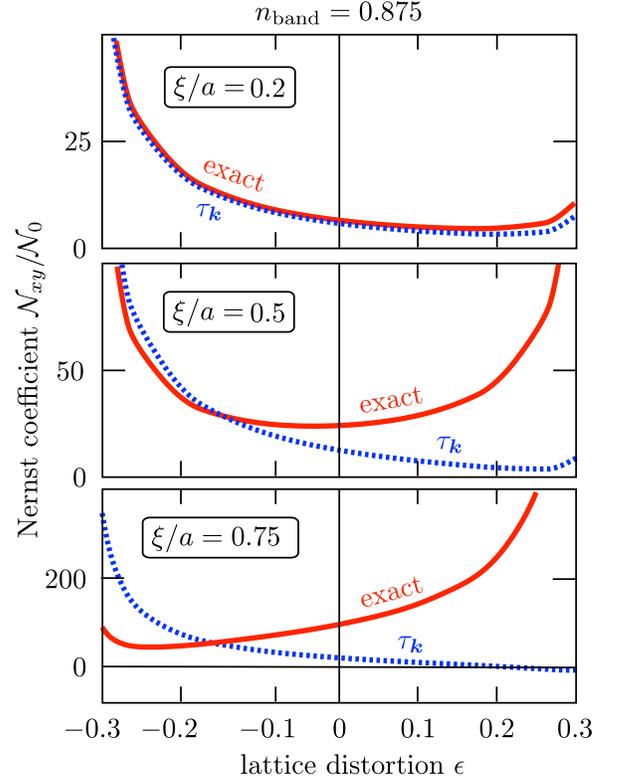}}
\caption{Dependence of the Nernst coefficient on the distortion $\epsilon$ of the square lattice at a fixed band filling $n_{\rm band}=0.875$, for three different values of the range $\xi$ of the scattering potential. The three panels show how the exact solution of the linearized Boltzmann equation (solid) starts out very close to the relaxation-time approximation (dotted) for nearly isotropic scattering, and then becomes progressively different as small-angle scattering begins to dominate.
}
\label{fig_eps_xi}
\end{figure}

\begin{figure}[tbh]
\centerline{\includegraphics[width=0.9\linewidth]{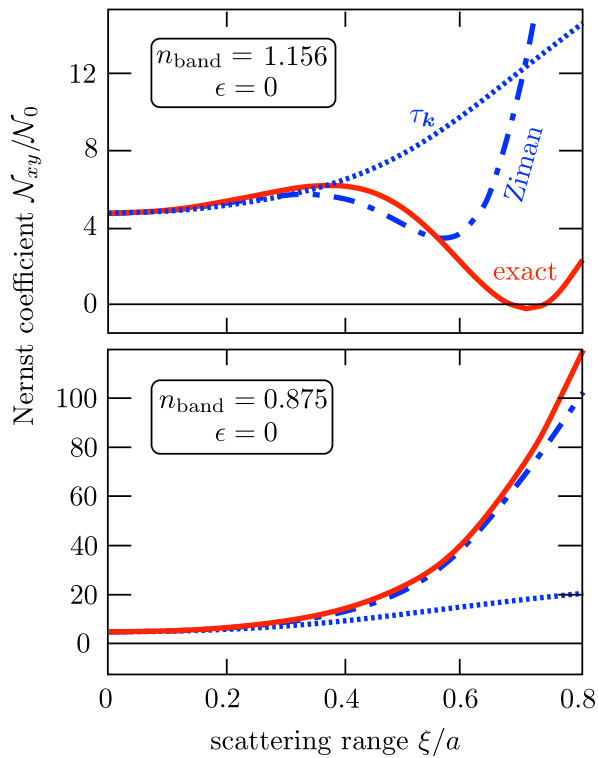}}
\caption{Dependence of the Nernst coefficient on the range $\xi$ of the scattering potential, for an undistorted square lattice ($\epsilon=0$). Two values of the band filling are shown in the upper and lower panel. The three curves in each panel correspond to: the exact solution of the linearized Boltzmann equation (solid), the relaxation-time approximation (dotted), and Ziman's improvement on the relaxation-time approximation (dash-dotted).
}
\label{fig_xi}
\end{figure}

\begin{figure}[tbh]
\centerline{\includegraphics[width=0.9\linewidth]{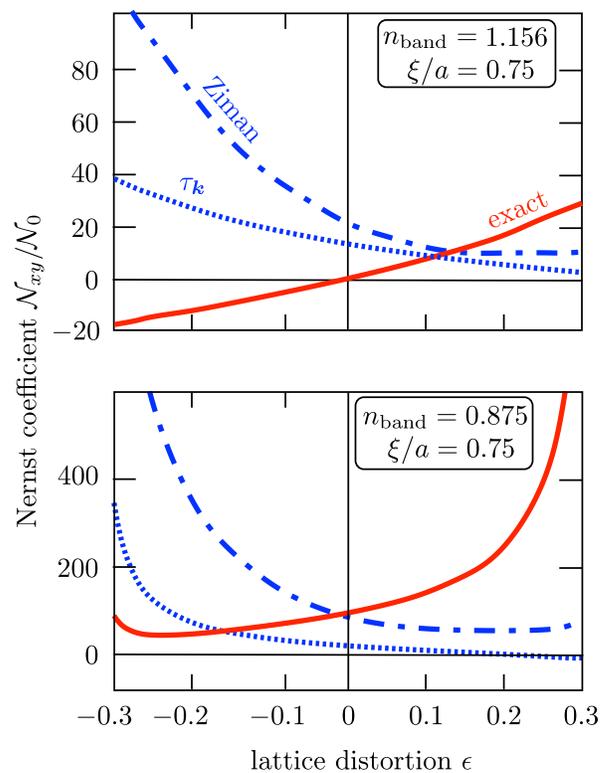}}
\caption{Same as Fig.\ \ref{fig_xi}, but now showing the dependence on the distortion $\epsilon$ of the square lattice for a fixed
range $\xi=0.75\,a$ of the scattering potential.
}
\label{fig_eps}
\end{figure}

We turn to a comparison of the Nernst effect in relaxation-time approximation with the exact solution of the linearized Boltzmann equation. For this comparison we need to specify an elastic scattering rate $Q(\bm{k},\bm{k}')=\delta(\varepsilon_{\bm{k}}-\varepsilon_{\bm{k}'})q(\bm{k},\bm{k}')$ and a dispersion relation $\varepsilon_{\bm{k}}$. 

For the scattering, we take a random impurity potential with range $\xi$. By increasing $\xi$ relative to the Fermi wave length, we can study the transition from isotropic scattering to (small-angle) forward scattering. We model the impurity potential by a sum of Gaussians, centered at the random positions $\bm{r}_{i}$ of the impurities,
\begin{equation}
U(\bm{r}) = \sum_{i}U_i \exp\left( - \frac{|\bm{r} - \bm{r}_i|^2}{\xi^2} \right).\label{Udef}
\end{equation}
The amplitude $U_i$ is uniformly distributed in $[-\delta,\delta]$. The correlator is
\begin{align}
&\langle U(\bm{r}) U(\bm{r}')\rangle = \frac{\pi}{6} \delta^2 \xi^2 n_{\rm imp}
\exp \left( - \frac{|\bm{r} - \bm{r}'|^2}{2\xi^2} \right),\label{Ucorr}\\
&\Rightarrow\langle |U(\bm{k})|^{2}\rangle=\tfrac{1}{12} \delta^2 \xi^4 n_{\rm imp} \exp\left(-\tfrac{1}{2}\xi^2 |\bm{k}|^2 \right),\label{Ucorrk}
\end{align}
where $n_{\rm imp}$ is the two-dimensional impurity density (number of impurities per area per layer). The resulting elastic scattering rate (in Born approximation) becomes
\begin{align}
q({\bm k,\bm k'})=& \gamma_0 \exp\left(-\tfrac{1}{2}\xi^2|\bm k-\bm k'|^2\right),\nonumber\\
&\gamma_0=\frac{\pi\delta^2 \xi^4 n_{\rm imp}}{6\hbar}.\label{qkk}
\end{align}
Values of $\xi/a$ of order unity are to be expected in the cuprates for scattering by impurities between the ${\rm CuO}_{2}$ planes, when $\xi$ is of the order of the interplane distance.

For the dispersion relation we follow a recent study of the Nernst effect in hole-doped cuprates,\cite{Hac09b} by taking the tight-binding dispersion of a distorted square lattice with first ($t_1$), second ($t_{2}$), and third ($t_{3}$) nearest-neigbor hopping:
\begin{align}
E({\bm k})={}&-2t_1\bigl[(1+\epsilon) \cos k_x  + (1-\epsilon) \cos k_y \bigr]\nonumber\\
&- 2 t_3 \bigl[(1+\epsilon)\cos 2k_x + (1-\epsilon)\cos 2k_y\bigr]\nonumber\\
&+4 t_2 \cos k_x \cos k_y.\label{Eksquare}
\end{align}
The lattice constant is $a$ and $\bm{k}$ is measured in units of $1/a$. The $C_{4}$ symmetry is distorted by the anisotropy parameter $\epsilon$, preserving reflection symmetry in the $x$ and $y$-axes. 

We use ratios of hopping parameters $t_2/t_1=0.32$, $t_3/t_2=0.5$, and compare two values of the band filling fractions $n_{\rm band}=1.156$ and $0.875$. (The corresponding Fermi energies at $\epsilon=0$ are $E_{F}=0$ and $E_{F}\approx - 0.97 t_{1}$ respectively, and are adjusted as $\epsilon$ is varied to keep $n_{\rm band}$ fixed.)

The Nernst coefficient is plotted in units of 
\begin{equation}
{\cal N}_{0}=\frac{t_1 a^4 Z_{0}}{\hbar\gamma_0}
=\frac{\pi^{2}k_{B}^{2}T}{3\hbar^{2}}\frac{t_{1}a^{4}}{\gamma_{0}}
.\label{NcalN}
\end{equation}
We show only ${\cal N}_{xy}$, since ${\cal N}_{yx}$ is related by
\begin{equation}
{\cal N}_{xy}(\epsilon)=-{\cal N}_{yx}(-\epsilon).\label{NxyNyxrelation}
\end{equation}
We compare three results for the Nernst coefficient:
\begin{itemize}
\item the exact solution of the linearized Boltzmann equation, from Eq.\ \eqref{calNLambda};
\item the momentum-dependent relaxation-time approximation, from Eq.\ \eqref{calNdefreltime};
\item Ziman's improvement on the relaxation-time approximation, from Eq.\ \eqref{taukdefZiman}.
\end{itemize}
We have have found that there is little difference between the momentum-dependent and momentum-independent relaxation-time approximations [Eqs.\ \eqref{calNdefreltime} and \eqref{calNdefreltime0}], so we only plot the former.

Results are shown in Figs.\ \ref{fig_eps_xi}--\ref{fig_eps}. Fig.\ \ref{fig_eps_xi} shows that the relaxation-time approximation agrees well with the exact solution for nearly isotropic scattering ($\xi\ll a$). With increasing $\xi$ small-angle scattering begins to dominate, and the relaxation-time approximation breaks down for $\xi\gtrsim 0.4\,a$. In Fig.\ \ref{fig_xi} we see that Ziman's improved approximation remains reliable over a somewhat larger range of $\xi$. Still, for a modestly large $\xi=0.75\,a$ also Ziman's approximation has broken down completely, see Fig.\ \ref{fig_eps}, giving wrong magnitude and sign of the Nernst coefficient.

\section{Conclusion}
\label{conclude}

In conclusion, we have shown that the relaxation-time approximation is not a reliable method to calculate the Nernst effect in the combined presence of band and scattering anisotropies. The deficiencies are qualitative, even the sign of the effect can come out wrong. Of course, the relaxation-time approximation remains a valuable tool to assess the effects of band anisotropy in the case of isotropic scattering.

We have based our comparison on parameters relevant for the cuprates,\cite{Hac09b} but we have only considered one possible mechanism (single-band elastic quasiparticle scattering) for the Nernst effect in cuprate superconductors. Other mechanisms (ambipolar diffusion, inelastic scattering, superconducting fluctuations) would require separate investigations.\cite{Beh09} It is hoped that the general framework provided here will motivate and facilitate work in that direction. 

\acknowledgments

This research was supported by the Dutch Science Foundation NWO/FOM and by an
ERC Advanced Investigator Grant.

\end{document}